\documentclass[fleqn,usenatbib]{mnras}
\usepackage{amsmath}
\usepackage{mathptmx}
\usepackage{txfonts}

\usepackage[T1]{fontenc}
\usepackage{ae,aecompl}

\usepackage{graphicx}	
\usepackage{amssymb}	
\usepackage{booktabs}
\usepackage{wasysym}
\usepackage{color}
\usepackage[table]{xcolor}
\usepackage{verbatim}
\usepackage{lscape}

\let\oldhref\href
\renewcommand{\href}[2]{\oldhref{#1}{\hbox{#2}}}

\definecolor{colorl1}{RGB}{0, 51, 153}
\definecolor{colorl2}{RGB}{153, 0, 0}
\definecolor{colorl3}{RGB}{179, 179, 0}
\definecolor{colorl4}{RGB}{51, 102, 0}

\definecolor{colorw1}{RGB}{51, 102, 255}
\definecolor{colorw2}{RGB}{255, 51, 0}
\definecolor{colorw3}{RGB}{255, 214, 51}
\definecolor{colorw4}{RGB}{51, 204, 51}

\newcommand{\hMpc}{{\ifmmode{h^{-1}{\rm Mpc}}\else{$h^{-1}$Mpc}\fi}}
\newcommand{\Mpc}{{\ifmmode{{\rm Mpc}}\else{Mpc}\fi}}
\newcommand{\hkpc}{{\ifmmode{h^{-1}{\rm kpc}}\else{$h^{-1}$kpc}\fi}}
\newcommand{\kpc}{{\ifmmode{ {\rm kpc} }\else{{\rm kpc}}\fi}}
\newcommand{\kms}{{\ifmmode{ {\rm km\,s^{-1}} }\else{ ${\rm km\,s^{-1}}$ }\fi}}
\newcommand{\hMsun}{{\ifmmode{h^{-1}{\rm {M_{\astrosun}}}}\else{$h^{-1}{\rm{M_{\astrosun}}}$}\fi}}
\newcommand{\Msun}{{\ifmmode{{\rm M}_{\astrosun}}\else{${\rm M}_{\astrosun}$}\fi}}
\newcommand{\Mhalo}{{\ifmmode{M_{\rm halo}}\else{$M_{\rm halo}$}\fi}}
\newcommand{\Rvir}{{\ifmmode{R_{\rm vir}}\else{$R_{\rm vir}$}\fi}}
\newcommand{\Mvir}{{\ifmmode{M_{\rm vir}}\else{$M_{\rm vir}$}\fi}}
\newcommand{\Mstar}{{\ifmmode{M_{\rm star}}\else{$M_{\rm star}$}\fi}}
\newcommand{\Vrot}{{\ifmmode{V_{\rm rot}}\else{$V_{\rm rot}$}\fi}}
\newcommand{\ltsima}{$\; \buildrel < \over \sim \;$}
\newcommand{\gtsima}{$\; \buildrel > \over \sim \;$}
\newcommand{\lsim}{\lower.5ex\hbox{\ltsima}}
\newcommand{\gsim}{\lower.5ex\hbox{\gtsima}}

\def\lesssim{\mathrel{\hbox{\rlap{\hbox{\lower4pt\hbox{$\sim$}}}\hbox{$<$}}}}
\def\gtrsim{\mathrel{\hbox{\rlap{\hbox{\lower4pt\hbox{$\sim$}}}\hbox{$>$}}}}

\newcommand{\beq}{\begin{equation}}
\newcommand{\eeq}{\end{equation}}
\def\beqa{\begin{eqnarray}}
\def\eeqa{\end{eqnarray}}
\def\LCDM{\ensuremath{\Lambda}CDM}

\def\head{ \vbox to 0pt{\vss \hbox to 0pt{\hskip 440pt\rm
      LA-UR-10-07069\hss} \vskip 25pt}}

\def \kms {\ifmmode  \,\rm km\,s^{-1} \else $\,\rm km\,s^{-1}  $ \fi }
\def \kpc {\ifmmode  {\,\rm kpc}  \else ${\rm  kpc}$ \fi  }  
\def \hkpc {\ifmmode  {h^{-1}\rm kpc}  \else ${h^{-1}\rm kpc}$ \fi  }  
\def \hMpc {\ifmmode  {h^{-1}\rm Mpc}  \else ${h^{-1}\rm Mpc}$ \fi  }  
\def \Mpch {\ifmmode  {h^{-1}\rm Mpc}  \else ${h^{-1}\rm Mpc}$ \fi  }  
\def \Msun {\ifmmode {\rm M}_{\astrosun} \else ${\rm M}_{\astrosun}$ \fi} 
\def \hMsun {\ifmmode h^{-1}\,\rm M_{\astrosun} \else $h^{-1}\,\rm M_{\astrosun}$ \fi}
\def \Gyr {\ifmmode\, \rm Gyr \else $\,$Gyr \fi}

\def \LCDM {\ifmmode \Lambda{\rm CDM} \else $\Lambda{\rm CDM}$ \fi}
\def \sig8 {\ifmmode \sigma_8 \else $\sigma_8$ \fi} 
\def \OmegaM {\ifmmode \Omega_{\rm m} \else $\Omega_{\rm m}$ \fi} 
\def \Omegab {\ifmmode \Omega_{\rm b} \else $\Omega_{\rm b}$ \fi} 
\def \OmegaL {\ifmmode \Omega_{\rm \Lambda} \else $\Omega_{\rm \Lambda}$\fi} 
\def \Deltavir {\ifmmode \Delta_{\rm vir} \else $\Delta_{\rm vir}$ \fi}
\def \rhocrit {\ifmmode \rho_{\rm crit} \else $\rho_{\rm crit}$ \fi}
\def \rhou {\ifmmode \rho_{\rm u} \else $\rho_{\rm u}$ \fi}
\def \zc {\ifmmode z_{\rm c} \else $z_{\rm c}$ \fi}

\title[The edge of galaxy formation III]{The edge of galaxy formation III:  The effects of warm dark matter on Milky Way satellites and field dwarfs}

\author[A.V. Macci\`o et al.]{Andrea V. Macci\`o$^{1,2}$\thanks{E-mail: maccio@nyu.edu},
Jonas Frings$^{3,2}$\thanks{E-mail: frings@mpia.de},
Tobias Buck$^{2}$, Aaron A. Dutton$^{1}$, 
\newauthor{Marvin Blank$^{1,4}$, Aura Obreja$^{5,1}$, Keri L. Dixon$^1$}
\\
$^{1}$New York University Abu Dhabi, PO Box 129188, Abu Dhabi, United Arab Emirates\\
$^{2}$Max-Planck-Institut f\"ur Astronomie, K\"onigstuhl 17, 69117 Heidelberg, Germany\\
$^{3}$Astronomisches Recheninstitut, Zentrum f{\"u}r Astronomie der Universit{\"a}t Heidelberg, Philosophenweg 12, 69120 Heidelberg, Germany \\
$^{4}$Institut f\"{u}r Theoretische Physik und Astrophysik, Christian-Albrechts-Universit\"{a}t zu Kiel, Leibnizstr. 15, D-24118 Kiel, Germany\\
$^{5}$Universit\"ats-Sternwarte, Ludwig-Maximilians-Universit\"at M\"unchen, Scheinerstr. 1, D-81679 M\"unchen, Germany\\
}

\date{Accepted XXX. Received YYY; in original form ZZZ}

\pubyear{2018}

\begin{document}

\label{firstpage}
\pagerange{\pageref{firstpage}--\pageref{lastpage}}
\maketitle
\begin{abstract}
In this third paper of the series, we investigate the effects of warm dark matter with a particle mass of $m_\mathrm{WDM}=3\,\mathrm{keV}$ on the smallest galaxies in our Universe. We present a sample of 21 hydrodynamical cosmological simulations of dwarf galaxies and 20 simulations of satellite-host galaxy interaction that we performed both in a Cold Dark Matter (CDM) and Warm Dark Matter (WDM) scenario.
In the WDM simulations, we observe a higher critical mass for the onset of star formation. Structure growth is delayed in WDM, as a result WDM haloes have a stellar population on average two Gyrs younger than their CDM counterparts. Nevertheless, despite this delayed star formation, CDM and WDM galaxies are both able to reproduce the observed scaling relations for velocity dispersion, stellar mass, size, and metallicity at $z=0$. WDM satellite haloes in a Milky Way mass host are more susceptible to tidal stripping due to their lower concentrations, but their galaxies can even survive longer than the CDM counterparts if they live in a dark matter halo with a steeper central slope. In agreement with our previous CDM satellite study we observe a  steepening of the WDM satellites' central dark matter density slope due to stripping. The difference in the average stellar age for satellite galaxies, between CDM and WDM, could be used in the future for disentangling these two models. 
\end{abstract}

\begin{keywords}
cosmology: theory -- dark matter -- galaxies: formation -- galaxies: kinematics and dynamics -- methods: numerical
\end{keywords}


\section{Introduction}\label{sec:introduction}

In the current standard model for the evolution of our Universe and the formation of structure therein gravity is described by general relativity with a cosmological constant $\Lambda$ \citep{Riess1998, Perlmutter1999}. The structure formation (at large scales and early times) is dominated by the Cold Dark Matter \citep[CDM,][]{Peebles1984} that, in contrast to baryonic matter, is not subject to pressure and other forces but gravity. In such a model, the formation of structure is happening in a hierarchical fashion, where the smallest haloes form first and merge to form more massive strucutres \citep{White1978, Blumenthal1984}. Once decoupled from the radiation field, gas can fall onto the dark matter haloes and eventually cool and form stars and finally galaxies.
Although being very successful in predicting the observed structure of the Universe on large ($>$Mpc) scales there still are open questions regarding the ability to reproduce observations on the scales of dwarf galaxies and Milky Way satellites 
\citep[see][]{Flores1994,Moore1994,Klypin1999,Moore1999,Boylan-Kolchin2011,Oh2015}.

In order to answer these questions and to shed light on the process of galaxy formation at the edge of its mass spectrum, in the first two papers of this series \citep{Maccio2017, Frings2017} (from now on referred to as \textit{paperI} and \textit{paperII})   we present a series of very high resolution hydrodynamical simulations (resolution $m_\mathrm{star}\approx 100\Msun$ and softening $\epsilon\approx 30\,\mathrm{pc}$) in dark matter haloes with masses $10^{8.5}-10^{10} \Msun$ (\textit{paperI}). 
We then studied the effect of tidal harassment from a possible Milky Way like central object in \textit{paperII}.

Our results showed that simulated dwarf and satellite galaxies in a CDM universe can successfully reproduce the observed scaling relations for Local Group dwarfs and satellites. We also showed that tidal interactions and stripping are very important ingredients to reproduce the scatter in these relations and to explain some observed peculiar objects \citep[e.g. Crater2,][]{Caldwell2017}. Given the success of CDM simulations it is interesting to ask the question whether other types of dark matter would leave  different imprints on these very low mass objects, and hence whether we can use observations of galactic satellites to learn more about the nature of dark matter.

A popular alternative to CDM is Warm Dark Matter (WDM); in this model particles still have a non-negligible thermal velocity at decoupling, and they can then free-stream out of the small density perturbations. This free-streaming creates a cut-off in the power spectrum and hence suppresses structure formation on small scales \citep{Bode2001}.

WDM models were first introduced to alleviate the aforementioned problems of CDM on small scales \citep{Colin2000,Lovell2012}. 
While it has been realized that baryonic physics provides a natural solution for the small-scale crisis \citep{Maccio2010,Sawala2014,Bullock2017,Buck2018}, WDM remains a  viable alternative to CDM, and it is possibly supported by the recent claims of the detection of a $3.5\,\mathrm{keV}$ line in the X-ray spectra of galaxy clusters and single galaxies, which can  be  interpreted  as  coming  from  the  decay  of  a  dark matter particle with the mass of $\approx 7$keV \citep[see][for a recent review]{Boyarsky2018}.

The inclusion of the effects of baryonic physics in the modelling of WDM has already been proven to be a key ingredient in better understanding the actual {\it temperature} of dark matter.
\cite{Maccio2010c} using N-body simulations coupled with a semi-analytic model for galaxy formation, showed that the luminosity function of the Milky Way satellites sets a lower limit of 2 keV for the mass 
of a WDM thermal candidate \citep[see also][]{Kang2013}. \citet{Governato2015} investigated the effect of a $2\,\mathrm{keV}$ WDM on the evolution of a single halo simulated with various star formation and feedback models. 
They found that star formation is both reduced and delayed by 1-2 Gyr  compared to the CDM runs, results confirmed recently by \citet{Chau2017} using a simple analytic model to link the star formation history
of a dark matter halo to its formation time, and by \cite{Bozek2018} using hydrodynamical cosmological simulations with resonant sterile neutrino dark matter.

Finally the possible presence of a  warm candidate has been searched for in other venues beyond faint galaxies. 
To cite a few examples, \cite{Bremer2018} used the enrichment of the intergalactic medium,
\cite{Wang2017} and \cite{Menci2018} studied global properties of galaxies in CDM and WDM using semi-analytical models, \cite{Dayal2017} looked at high-z direct collapse black holes in WDM and \cite{Bose2016} and \cite{Villanueva2018}  explored the effects of WDM on cosmic reionization, while \cite{Lovell2017} have presented a series of hydrodynamical simulations of sterile neutrino dark matter in the Local Group context.

In this paper, we extend and complete previous results on the effect of WDM on the properties of the faintest galaxies in the universe. 
We focus our attention on a WDM thermal candidate with a mass of $3\,\mathrm{keV}$, such a candidate has the lowest possible mass in agreement with limits set by the Lyman-alpha forest \citep{Viel2013,Irsic2017, Garzilli2018}. 
We simulated a sample of 21 haloes both in WDM and CDM to study isolated dwarf galaxies in WDM. Four of those haloes are then used as initial conditions for the study of satellite-host interactions.

This paper is organized as follows, in Section \ref{sec:simulations} we introduce the code and the sample of hydrodynamical cosmological simulations. In Section \ref{sec:results}, we present the comparison of the evolution of the dwarf galaxies in CDM and WDM investigating observable scaling relations, stellar ages and dark matter structure. Further we study the effects of tidal stripping of Milky Way satellites in CDM and WDM. In Section \ref{sec:discussion}, we finally summarise and discuss our findings.

\section{Simulations} \label{sec:simulations}


\begin{table*}
\centering
\caption{Name, virial mass, stellar mass, stellar particle number and dark matter particle mass. Data from the CDM and WDM runs are indicated by blue and red background colours, respectively. 
The gravitational softening for the stellar particles ranges from $\varepsilon_\mathrm{star}=90\,\mathrm{pc}$ to $\varepsilon_\mathrm{star}=40\,\mathrm{pc}$ depending on resolution. Galaxies marked with an asterisk (*) have been used as initial conditions for the satellites host interaction simulations.} 
\label{tab:sample}
\begin{tabular}{lccccccr}

Name& \cellcolor{blue!15}$M_\mathrm{200}[\Msun]$ &\cellcolor{red!15} $M_\mathrm{200}[\Msun]$ &\cellcolor{blue!15} $M_\mathrm{star}[\Msun]$&\cellcolor{red!15} $M_\mathrm{star}[\Msun]$  &\cellcolor{blue!15}$N_\mathrm{star}$&\cellcolor{red!15}$N_\mathrm{star}$& $m_\mathrm{DM}[\Msun]$\\ 
\hline
 & \cellcolor{blue!15} CDM &\cellcolor{red!15} WDM &\cellcolor{blue!15} CDM &\cellcolor{red!15} WDM  &\cellcolor{blue!15}CDM &\cellcolor{red!15}WDM & \\ 

\hline\hline
g8.94e8&	 \cellcolor{blue!15}8.19 $10^8$ &\cellcolor{red!15} 2.31 $10^{8} $&	\cellcolor{blue!15}0 &\cellcolor{red!15} 0&\cellcolor{blue!15} 0&\cellcolor{red!15}0&2.38 $10^{3}$\\ \hline	
g1.89e9&	 \cellcolor{blue!15}1.42 $10^9$ &\cellcolor{red!15}	1.07 $10^9$&\cellcolor{blue!15}0 &\cellcolor{red!15} 0&\cellcolor{blue!15} 0&\cellcolor{red!15}0&2.38 $10^{3}$ \\ \hline	
g3.54e9&	 \cellcolor{blue!15}2.85 $10^9$ &\cellcolor{red!15}	2.51 $10^9$&	\cellcolor{blue!15}3.89 $10^{5}$ &\cellcolor{red!15} 0&\cellcolor{blue!15} 4704&\cellcolor{red!15}0&2.38 $10^{3}$ \\ \hline
g3.67e9&	 \cellcolor{blue!15}3.12 $10^9$ &\cellcolor{red!15}	2.56 $10^9$&	\cellcolor{blue!15}9.20 $10^{4}$ &\cellcolor{red!15} 0&\cellcolor{blue!15} 1110&\cellcolor{red!15}0&2.38 $10^{3}$ \\ \hline
g4.36e9&	 \cellcolor{blue!15}9.71 $10^9$ &\cellcolor{red!15}	5.43 $10^9$&	\cellcolor{blue!15}3.30 $10^{4}$ &\cellcolor{red!15} 0&\cellcolor{blue!15} 38&\cellcolor{red!15}0& 8.03 $10^{3}$ \\ \hline
g4.48e9*&	 \cellcolor{blue!15}3.63 $10^9$ &\cellcolor{red!15}	3.13 $10^9$&	\cellcolor{blue!15}6.51 $10^{5}$ &\cellcolor{red!15} 9.69 $10^{4}$&\cellcolor{blue!15} 7883&\cellcolor{red!15}1172& 2.38 $10^{3}$ \\ \hline
g4.99e9&	 \cellcolor{blue!15}5.70 $10^9$ &\cellcolor{red!15}	4.29 $10^9$&	\cellcolor{blue!15}3.76 $10^{5}$ &\cellcolor{red!15} 0&\cellcolor{blue!15} 538&\cellcolor{red!15}0& 1.90 $10^{4}$ \\ \hline
g5.22e9&	 \cellcolor{blue!15}6.30 $10^9$ &\cellcolor{red!15}	3.48 $10^9$&	\cellcolor{blue!15}1.17 $10^{5}$ &\cellcolor{red!15} 0&\cellcolor{blue!15} 171&\cellcolor{red!15}0& 1.90 $10^{4}$ \\ \hline
g5.59e9&	 \cellcolor{blue!15}6.43 $10^9$ &\cellcolor{red!15}	6.28 $10^9$&	\cellcolor{blue!15}1.73 $10^{6}$ &\cellcolor{red!15} 7.00 $10^{5}$&\cellcolor{blue!15} 2509&\cellcolor{red!15}1012& 1.90 $10^{4}$ \\ \hline
g6.31e9&	 \cellcolor{blue!15}5.13 $10^9$ &\cellcolor{red!15}	4.48 $10^9$&	\cellcolor{blue!15}3.71 $10^{5}$ &\cellcolor{red!15} 2.48 $10^{2}$&\cellcolor{blue!15} 4457&\cellcolor{red!15}3& 2.38 $10^{3}$ \\ \hline
g7.05e9&	 \cellcolor{blue!15}1.05 $10^{10}$ &\cellcolor{red!15}	8.68 $10^9$&	\cellcolor{blue!15}2.26 $10^{6}$ &\cellcolor{red!15} 2.01 $10^{6}$&\cellcolor{blue!15} 3230&\cellcolor{red!15}2854& 1.90 $10^{4}$ \\ \hline
g8.63e9&	 \cellcolor{blue!15}5.59 $10^9$ &\cellcolor{red!15}	4.34 $10^9$&	\cellcolor{blue!15}5.90 $10^{5}$ &\cellcolor{red!15} 0&\cellcolor{blue!15} 2022&\cellcolor{red!15}0& 8.03 $10^{3}$ \\ \hline
g9.91e9*&	 \cellcolor{blue!15}7.42 $10^9$ &\cellcolor{red!15}	7.25 $10^9$&	\cellcolor{blue!15}1.50 $10^{6}$ &\cellcolor{red!15} 8.96 $10^{5}$&\cellcolor{blue!15} 5238&\cellcolor{red!15}3097& 8.03 $10^{3}$ \\ \hline
g1.17e10*&	 \cellcolor{blue!15}8.60 $10^9$ &\cellcolor{red!15}	9.07 $10^9$&	\cellcolor{blue!15}3.34 $10^{6}$ &\cellcolor{red!15} 2.90 $10^{6}$&\cellcolor{blue!15} 11613&\cellcolor{red!15}10109& 8.03 $10^{3}$ \\ \hline
g1.18e10&	 \cellcolor{blue!15}1.09 $10^{10}$ &\cellcolor{red!15}	1.44 $10^{10}$&	\cellcolor{blue!15}3.37 $10^{6}$ &\cellcolor{red!15} 8.73 $10^{6}$&\cellcolor{blue!15} 4887&\cellcolor{red!15}12551& 1.90 $10^{4}$ \\ \hline
g1.23e10*&	 \cellcolor{blue!15}9.06 $10^9$ &\cellcolor{red!15}	1.32 $10^{10}$&	\cellcolor{blue!15}1.58 $10^{6}$ &\cellcolor{red!15} 2.36 $10^{6}$&\cellcolor{blue!15} 2278&\cellcolor{red!15}3405& 1.90 $10^{4}$ \\ \hline
g1.44e10&	 \cellcolor{blue!15}1.69 $10^{10}$ &\cellcolor{red!15}	1.22 $10^{10}$&	\cellcolor{blue!15}6.64 $10^{6}$ &\cellcolor{red!15} 3.24 $10^{6}$&\cellcolor{blue!15} 9562&\cellcolor{red!15}4664& 1.90 $10^{4}$ \\ \hline
g1.47e10&	 \cellcolor{blue!15}1.52 $10^{10}$ &\cellcolor{red!15}	1.03 $10^{10}$&	\cellcolor{blue!15}8.92 $10^{6}$ &\cellcolor{red!15} 5.15 $10^{6}$&\cellcolor{blue!15} 12910&\cellcolor{red!15}7476& 1.90 $10^{4}$ \\ \hline
g1.50e10&	 \cellcolor{blue!15}1.40 $10^{10}$ &\cellcolor{red!15}	1.22 $10^{10}$&	\cellcolor{blue!15}3.32 $10^{6}$ &\cellcolor{red!15} 2.31 $10^{6}$&\cellcolor{blue!15} 4840&\cellcolor{red!15}3222& 1.90 $10^{4}$ \\ \hline
g1.95e10&	 \cellcolor{blue!15}1.37 $10^{10}$ &\cellcolor{red!15}	1.20 $10^{10}$&	\cellcolor{blue!15}3.79 $10^{6}$ &\cellcolor{red!15} 2.33 $10^{6}$&\cellcolor{blue!15} 5429&\cellcolor{red!15}3327& 1.90 $10^{4}$ \\ \hline
g2.94e10&	 \cellcolor{blue!15}3.22 $10^{10}$ &\cellcolor{red!15}	3.05 $10^{10}$&	\cellcolor{blue!15}5.63 $10^{7}$ &\cellcolor{red!15} 4.75 $10^{7}$&\cellcolor{blue!15} 81608&\cellcolor{red!15}68629& 1.90 $10^{4}$ \\ \hline                     
\end{tabular}
\end{table*}

\begin{table*}
\centering
\caption{Name, projected stellar half-mass radi, stellar line-of-sight velocity dispersion and mean metallicity. Galaxies marked with an asterisk (*) have been used as initial conditions for the satellites host interaction simulations.
Objects with no values remain dark through the whole simulation.} 
\label{tab:sample_observables}
\begin{tabular}{lcccccr}

Name& \cellcolor{blue!15}$r_\mathrm{h}[\kpc]$ &\cellcolor{red!15} $r_\mathrm{h}[\kpc]$ &\cellcolor{blue!15} $\sigma[\mathrm{km}\mathrm{s}^{-1}]$&\cellcolor{red!15} $\sigma[\mathrm{km}\mathrm{s}^{-1}]$  &\cellcolor{blue!15}[Fe/H]&\cellcolor{red!15}[Fe/H]\\
\hline
&  \cellcolor{blue!15}CDM  &\cellcolor{red!15} WDM  &\cellcolor{blue!15} CDM &\cellcolor{red!15} WDM   &\cellcolor{blue!15}CDM &\cellcolor{red!15}WDM \\ \hline\hline
g8.94e8 &\cellcolor{blue!15} -&\cellcolor{red!15} -&\cellcolor{blue!15} -&\cellcolor{red!15} -&\cellcolor{blue!15} -&\cellcolor{red!15} -   \\ \hline
g1.89e9 &\cellcolor{blue!15} -&\cellcolor{red!15} -&\cellcolor{blue!15} -&\cellcolor{red!15} -&\cellcolor{blue!15} -&\cellcolor{red!15} -   \\ \hline
g3.54e9 &\cellcolor{blue!15} 0.361 &\cellcolor{red!15} -&\cellcolor{blue!15} 12.4 &\cellcolor{red!15} -&\cellcolor{blue!15}  -2.286 &\cellcolor{red!15} -  \\ \hline
g3.67e9 &\cellcolor{blue!15} 0.280 &\cellcolor{red!15} -&\cellcolor{blue!15} 8.39 &\cellcolor{red!15} -&\cellcolor{blue!15}  -2.496 &\cellcolor{red!15} -  \\ \hline
g4.36e09 &\cellcolor{blue!15} 0.171 &\cellcolor{red!15} -&\cellcolor{blue!15} 6.378 &\cellcolor{red!15} -&\cellcolor{blue!15}  -3.194 &\cellcolor{red!15} -  \\ \hline
g4.48e9* &\cellcolor{blue!15}  0.225 &\cellcolor{red!15} 0.163 &\cellcolor{blue!15}  9.469 &\cellcolor{red!15} 6.172  &\cellcolor{blue!15}  -1.947 &\cellcolor{red!15} -2.485  \\  \hline
g4.99e09 &\cellcolor{blue!15} 0.239 &\cellcolor{red!15} -&\cellcolor{blue!15} 7.309 &\cellcolor{red!15} -&\cellcolor{blue!15}  -1.886 &\cellcolor{red!15} -  \\ \hline
g5.22e09 &\cellcolor{blue!15} 0.170 &\cellcolor{red!15} -&\cellcolor{blue!15} 6.499 &\cellcolor{red!15} -&\cellcolor{blue!15}  -2.434 &\cellcolor{red!15} -  \\ \hline
g5.59e09 &\cellcolor{blue!15}  0.430 &\cellcolor{red!15} 0.264 &\cellcolor{blue!15}  11.706 &\cellcolor{red!15} 7.619  &\cellcolor{blue!15}  -1.787 &\cellcolor{red!15} -1.898  \\  \hline
g6.31e9 &\cellcolor{blue!15} 0.159 &\cellcolor{red!15} -&\cellcolor{blue!15} 8.536 &\cellcolor{red!15} -&\cellcolor{blue!15}  -1.978 &\cellcolor{red!15} -  \\ \hline
g7.05e09 &\cellcolor{blue!15}  0.555 &\cellcolor{red!15} 0.729 &\cellcolor{blue!15}  10.036 &\cellcolor{red!15} 10.861  &\cellcolor{blue!15}  -1.721 &\cellcolor{red!15} -1.862  \\  \hline
g8.63e9 &\cellcolor{blue!15} 0.258 &\cellcolor{red!15} -&\cellcolor{blue!15} 8.783 &\cellcolor{red!15} -&\cellcolor{blue!15}  -1.932 &\cellcolor{red!15} -  \\ \hline
g9.91e9* &\cellcolor{blue!15}  0.364 &\cellcolor{red!15} 0.263 &\cellcolor{blue!15}  9.099 &\cellcolor{red!15} 8.540  &\cellcolor{blue!15}  -1.752 &\cellcolor{red!15} -1.797  \\  \hline
g1.17e10* &\cellcolor{blue!15}  0.385 &\cellcolor{red!15} 0.378 &\cellcolor{blue!15}  11.114 &\cellcolor{red!15} 10.341  &\cellcolor{blue!15}  -1.660 &\cellcolor{red!15} -1.735  \\  \hline
g1.18e10 &\cellcolor{blue!15}  0.526 &\cellcolor{red!15} 0.841 &\cellcolor{blue!15}  12.974 &\cellcolor{red!15} 14.593  &\cellcolor{blue!15}  -1.671 &\cellcolor{red!15} -1.659  \\  \hline
g1.23e10* &\cellcolor{blue!15}  0.498 &\cellcolor{red!15} 0.475 &\cellcolor{blue!15}  9.968 &\cellcolor{red!15} 11.748  &\cellcolor{blue!15}  -1.727 &\cellcolor{red!15} -1.777  \\  \hline
g1.44e10 &\cellcolor{blue!15}  1.486 &\cellcolor{red!15} 0.669 &\cellcolor{blue!15}  14.048 &\cellcolor{red!15} 11.796  &\cellcolor{blue!15}  -1.705 &\cellcolor{red!15} -1.741  \\  \hline
g1.47e10 &\cellcolor{blue!15}  0.801 &\cellcolor{red!15} 0.986 &\cellcolor{blue!15}  13.356 &\cellcolor{red!15} 14.361  &\cellcolor{blue!15}  -1.6156 &\cellcolor{red!15} -1.748  \\  \hline
g1.50e10 &\cellcolor{blue!15}  0.869 &\cellcolor{red!15} 0.552 &\cellcolor{blue!15}  14.441 &\cellcolor{red!15} 10.463  &\cellcolor{blue!15}  -1.767 &\cellcolor{red!15} -1.916  \\  \hline
g1.95e10 &\cellcolor{blue!15}  0.462 &\cellcolor{red!15} 0.403 &\cellcolor{blue!15}  12.561 &\cellcolor{red!15} 10.316  &\cellcolor{blue!15}  -1.670 &\cellcolor{red!15} -1.771  \\  \hline
g2.94e10 &\cellcolor{blue!15}  1.477 &\cellcolor{red!15} 1.763 &\cellcolor{blue!15}  20.911 &\cellcolor{red!15} 20.485  &\cellcolor{blue!15}  -1.423 &\cellcolor{red!15} -1.463  \\  \hline

\end{tabular}
\end{table*}
\subsection{Cosmological simulations} \label{sec:cosmosims}

We performed a total of 42 hydrodynamical zoom-in simulations of 21 haloes both in a CDM and WDM scenario using the smoothed particle hydrodynamics code {\sc gasoline2}  \citep{Wadsley2017}. Compared to the predecessor papers \citep{Maccio2017, Frings2017}, we updated the cosmological parameters according to \cite{Planck2014}: Hubble parameter $H_0$= 67.1 \kms Mpc$^{-1}$, matter density $\Omega_\mathrm{m}=0.3175$, dark energy density
$\Omega_{\Lambda}=1-\Omega_\mathrm{m} -\Omega_\mathrm{r}=0.6824$, baryon density
$\Omega_\mathrm{b}=0.0490$, normalization of the power spectrum $\sigma_8 = 0.8344$, slope of the inital power spectrum $n=0.9624$. 
 
The code setup is adopted from the Numerical Investigation of a Hundred Astrophysical Objects (NIHAO) project \citep{Wang2015}. The code includes metal gas cooling, chemical enrichment, star formation and stellar feedback. 
For the density threshold for star formation, we chose $\rho_\mathrm{min}=10.3\,\mathrm{cm}^{-3}$ according to the mass of 50 gas particles (smoothing kernel) in a volume spanned by the softening length \citep[for more details see][]{Wang2015}. If the gas reaches $\rho_\mathrm{min}$, it can form stars with an efficiency of $c_\star=0.1$. Gas cooling incorporates metal line cooling as described
in \citet{Shen2010}, Compton cooling, photoionisation, and heating from the ultraviolet background following \citet{Haardt2012}. 
The stellar feedback is implemented as a SN blast-wave feedback described in \citet{Stinson2006}, we also include the effect of the radiation from massive stars (early stellar feedback) following \cite{Stinson2013}.

The haloes were initally selected as isolated overdensities in two dark matter only cosmological volume simulations. The first box with a volume of $20^3 \,\Mpc^3$ and $300^3$ particles was already used in the NIHAO project \citep{Wang2015}, and the CDM runs of the haloes selected from this box are galaxies from the NIHAO sample. In addition, we ran a box with a volume of $10\,~ \Mpc^3$ and $600^3$ particles from which we chose further haloes to increase the size of our sample. We also ran both volumes in a WDM scenario. To describe the damping of small scales introduced by the streaming velocities, we start  from a WDM power spectrum $P_\mathrm{WDM}$ that is related to the CDM power spectrum $P_\mathrm{CDM}$ by the transfer function $T^2(k)$, where $k$ is the wavenumber.
We use a fitting formula
\begin{align}
T^2(k)=\frac{P_\mathrm{WDM}}{P_\mathrm{CDM}}=\left[1+(\alpha k)^{2\nu}\right]^{-10/\nu}
\end{align}
suggested by \citet{Bode2001} to calculate the transfer function.
The parameters are set to $\nu=1.12$ and 
\begin{align}
\alpha=0.049\,\left( \frac{m_\mathrm{WDM}}{1\,\mathrm{keV}}\right)^{-1.11}\,\left( \frac{\Omega_\mathrm{DM}}{0.25}\right)^{0.11}\,\left(\frac{h}{0.7} \right)^{1.22} h^{-1} \Mpc
\end{align}
following \citet{Viel2005}.  In principle, one should also add a thermal contribution to the velocities of the WDM particles \citep[e.g.][]{Bode2001}. On the other hand due to our choice of the WDM mass (3 keV) and our resolution, such a contribution will be negligible with respect to the velocity coming from the gravitational potential \citep{Maccio2012b}, and hence we decide to not include it. 
Starting from the cosmological volume simulations, we perform zoom-in simulations of the selected  (same)\footnote{The CDM counterparts of the WDM haloes were identified by using the dark matter particles' ID in low resolution run,  on average more than 50 percent of the IDs were matched.} object both in the CDM and WDM scenario. 
Being the half mode mass \citep[e.g.][]{Leo2017} for our 3 keV thermal relic of the order of $3\times10^8 \Msun$, we expect all selected haloes
to form in both cosmologies.

For the NIHAO objects we keep the resolution given by the NIHAO simulations while for the newly selected objects we choose the zoom factor such that we have at least $10^6$ dark matter particles inside the virial radius at redshift $z=0$. The dark matter particle mass in the zoom-in region as well as virial and stellar mass and stellar particle number (in 10 per cent of the virial radius) for each simulation is shown in Table \ref{tab:sample}. Gas particles have an inital mass of 
$m_\mathrm{gas,init}= m_\mathrm{DM} \Omega_{\rm b}/(\Omega_{\rm m} -\Omega_{\rm b})$,
while stellar particle start with inital masses $m_\mathrm{star, init}=\frac{1}{3}m_\mathrm{gas,init}$.
Table \ref{tab:sample_observables} shows the structural properties (2D half-mass radi, stellar velocity dispersion and metallitcity) of all galaxies in the WDM and CDM runs.

\subsection{Satellite initial conditions}	
From the  $z=1$ outputs of the cosmological simulations of the galaxies \textit{g1.23e10}, \textit{g1.17e10}, \textit{g9.91e9} and \textit{g4.48e9}, we proceed as described in \textit{paperII} \citep{Frings2017} by cutting the galaxies from their surroundings and evolving them  in an analytic Milky Way halo and disc potential with a gas removal scheme 
mimicking the effects of ram pressure. For the dark matter halo we use a Navarro, Frenk \& White (NFW) potential
\citep{nfw1996} with a mass $M_{200} =  10^{12}\Msun$,
a concentration parameter $c = 10$ \citep{Dutton2014},
and a virial radius $r_{200} = 210\kpc$. The stellar body is represented by a Miyamoto \& Nagai potential \citep{Miyamoto1975} with a disc mass $M_\mathrm{disc} = 5\times 10^{10}\Msun$, 
disc scale length $R_\mathrm{disc} = 3.0\kpc$ and height $h_\mathrm{disc} = 0.3\kpc$. Each satellite is then evolved on the five orbits listed in Table \ref{tab:orbits}. All trajectories start at the virial radius of the host halo in the plane of the disc, and the satellite will undergo two pericentre passages on all orbits until $z=0$. Each cut out galaxy is also evolved in isolation (no potential, no gas removal) from $z=1$ to $z=0$, we refer to these runs as ``isolated'' galaxies.

\begin{table}
\centering
\caption{Compilation of the different orbit scenarios with their inital velocity, pericentre distance and orbit inclination with respect to the host galaxy disc. All orbits originate at the coordinates $x=\kpc$, $y=z=0$.
$\tau_{\rm ram}$ is the time-scale for a complete removal of the gas due to ram pressure \citep[for more details see][]{Frings2017}.}
\label{tab:orbits}
\begin{tabular}{lcccr}

Name& $(v_x,\,v_y,\,v_z) \,[v_{200}]$  &$\tau_\mathrm{ram}\,[\Gyr]$& $r_\mathrm{min}\,[\kpc]$ &$\vartheta \,[\deg]$\\ \hline\hline
{\textit{orbitI}}&	$(-0.45,\,0.3,\,0)$&	1.5&25.52&0\\ \hline	
{\textit{orbitII}}&$(-0.45,\,0,\,0.3)$&	1.5&25.46&90\\ \hline	
{\textit{orbitIII}}&$(-0.2,\,0.2,\,0.2)$&1.5&25.26&45\\ \hline	
{\textit{orbitIV}}&$(-0.5,\,0,\,0.1)$&	1.4&7.94&90\\ \hline	
{\textit{orbitV}}&	$(-0.5,\,0.1,\,0)$&	1.4&7.2&0\\
\end{tabular}
\end{table}

\section{Results}\label{sec:results}
\subsection{Effects of WDM on field galaxies}
First we present the comparison between CDM and WDM simulations for our cosmological runs at $z=0$ (note that this is different from \textit{paperI}, where the field galaxies were analyzed at $z=1$).

\subsubsection{Halo masses}\label{sec:halo_masses}
\begin{figure}
\includegraphics[width=0.47\textwidth]{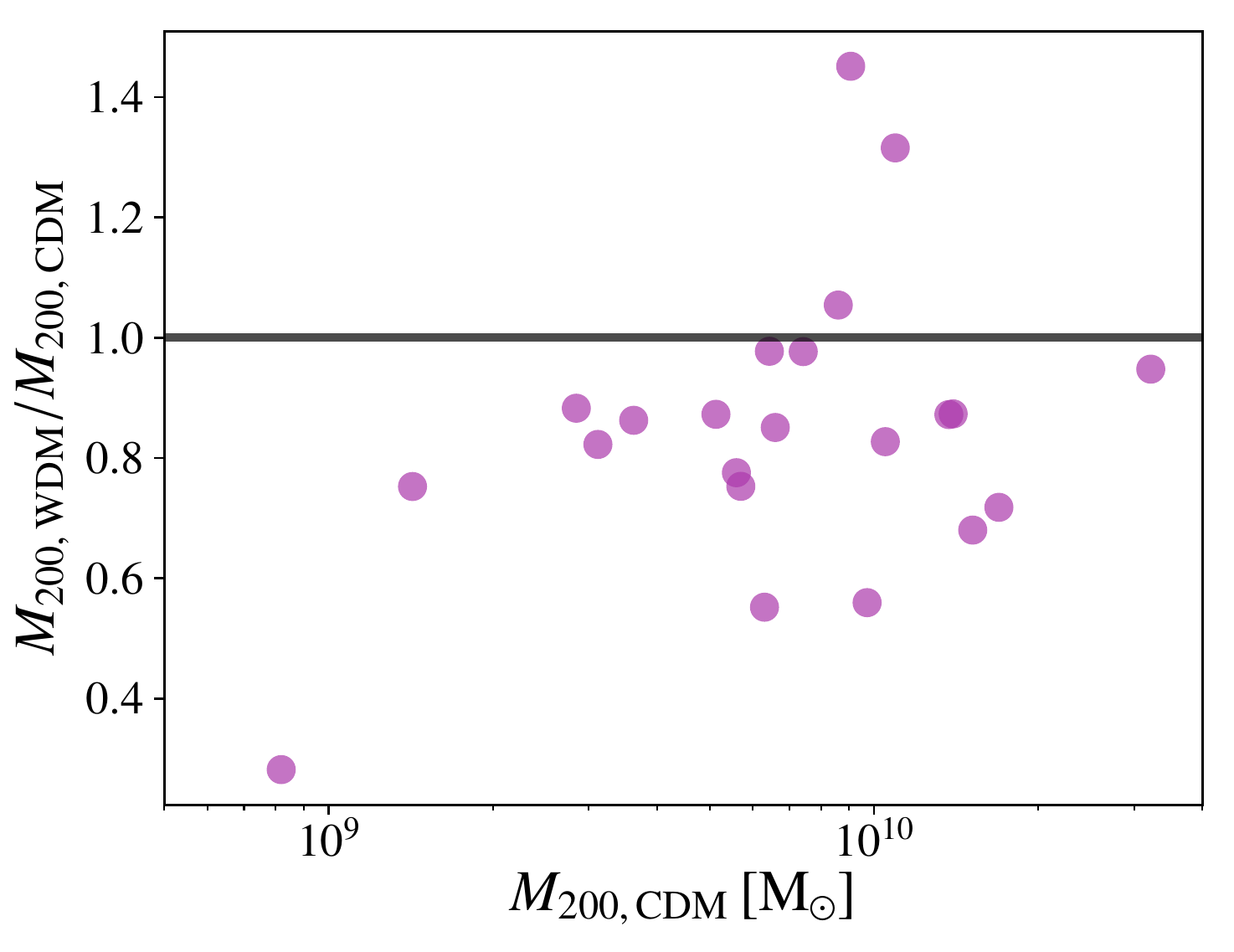}
\vspace{-.35cm}
\caption{Virial mass ratio of WDM to CDM run as a function of virial masses in CDM. The black line indicates the ratio of one and purple points are the simulation data.}
\label{fig:field_massratio}
\end{figure}
\begin{figure}
\includegraphics[width=0.47\textwidth]{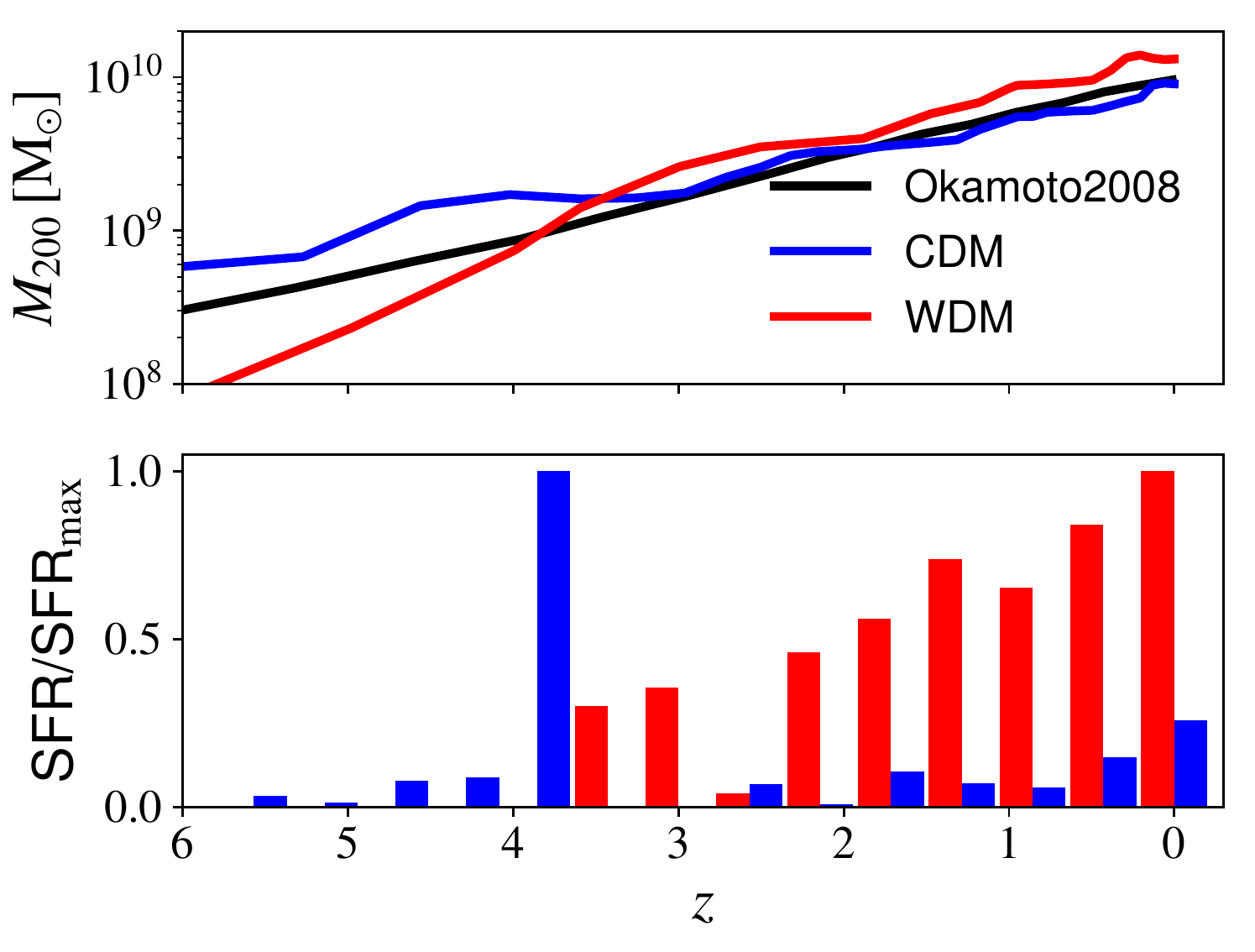}
\vspace{-.35cm}
\caption{Evolution of the virial mass (upper panel) and star formation rate (lower panel) of the simulation \textit{g1.23e10}. The black line indicates the mass at which haloes lose half their baryonic mass due to the UV background \citep{Okamoto2008}. }
\label{fig:field_masshist}
\end{figure}

Structure formation is delayed in WDM w.r.t. CDM, and we aim to quantify this by first looking at the halo masses. Throughout this work the virial masses and radii are always given with respect to $200\,\rho_\mathrm{crit}$ (with $\rho_\mathrm{crit}$ being the critical density). In Fig. \ref{fig:field_massratio} the ratio of virial masses in the WDM and CDM run is plotted against the CDM virial mass. The black line indicates the ratio of one and purple points are the simulation data. We can see that in a WDM scenario haloes are consistently less massive especially for masses below $10^{10}\, \Msun$ in agreement with linear theory expectations and previous studies \citep [e.g.][]{Bozek2016}. At higher masses a few of the simulated WDM haloes are more massive than their CDM counterpart; his is partially due to the scatter introduced by the stochasticity of the merger trees.
        To look at this issue in detail, we investigate the mass accretion histories of one of the haloes that ends up with a larger virial mass in WDM compared to CDM. In the upper panel  of Fig. \ref{fig:field_masshist}, we show the evolution of the virial mass of halo \textit{g1.23e10}. The black line is the critical mass from \citet{Okamoto2008} indicating the mass at which haloes lose half their baryonic mass due to the UV background. The star formation rate is shown in the lower panel. In the CDM scenario, there is a merger happening at $z\approx 4.5$ that led to a massive star formation event at $z=4$ (see peak in the lower panel) contributing to a large fraction of the total stellar mass. After that, the halo evolves very close to the Okamoto model with little star formation. In WDM however, that merger happens at a later time ($z\approx 3.5$), and the halo evolves far above the Okamoto critical mass with continuously increasing star formation as shown in the lower panel. The occurrence of such differences in the mass accretion histories of CDM - WDM counterparts shows the importance of a large sample of simulated galaxies not to be mislead by possible outliers. 

\subsubsection{Star formation and luminous fraction}
\begin{figure}
\includegraphics[width=0.47\textwidth]{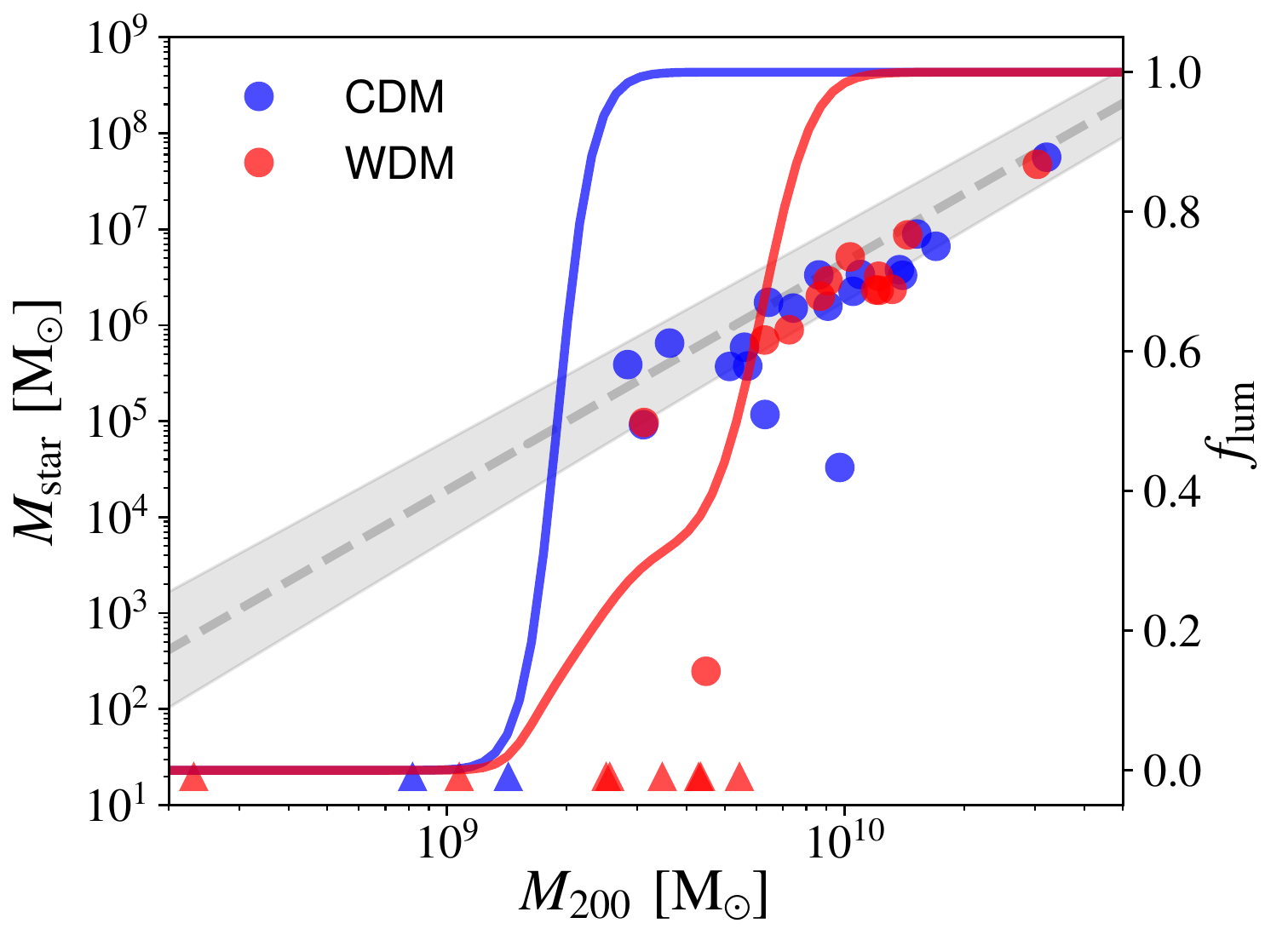}
\vspace{-.35cm}
\caption{Stellar mass versus the virial mass of the halo. On the second $y$-axis, the luminous fraction of satellites at given halo mass is shown as solid lines. Triangles indicate haloes that did not form stars.}
\label{fig:field_moster}
\end{figure}

In Fig. \ref{fig:field_moster} we show the stellar mass of the galaxies (i.e. stars inside $10$ per cent of the virial radius) as a function of the virial mass. The grey dashed line in the background is the low mass extrapolation of the abundance matching from \citet{Moster2013} and the associated scatter around the relation. Blue and red circles denote haloes that form galaxies in their centre in CDM and WDM, respectively, and triangles the haloes that stay dark until redshift zero. 

We  see that above a mass scale of $5\times 10^9\Msun$ in both dark matter scenarios haloes will always form stars, followed by a region where we can find both luminous and dark haloes and finally a minimum mass of $2\times 10^9\Msun$ below which no stars can form. 
Although halo masses from  CDM and WDM runs differ (as pointed out in Section \ref{sec:halo_masses}) the abundance matching relation is still fulfilled for the star forming WDM haloes. In other words, WDM may affect the final value of the halo mass, but not the star formation efficiency.

However, it seems that for the WDM haloes the critical halo mass for star formation is shifted to higher masses. To visualize this effect we also show the luminous fraction on the second $y$-axis as solid lines. 
We estimate the luminous fraction by smoothing the luminous galaxy counts with a Gaussian kernel at fixed width in logarithmic space. 
The two lines suggest that the transition from dark to luminous haloes is less sharp in WDM with respect to CDM. In WDM there is a large mass range where dark and luminous haloes coexist\footnote{A visual inspection of the WDM dark haloes showed that these haloes are fully formed
in their dark matter component but they failed to accrete any gas due to the UV background.}.
 In other works \citep{Sawala2016b,Buck2018} a broader transition region is observed also for CDM though at lower resolution. A larger sample of haloes is needed to better quantify the strength of this effect.

\begin{figure}
\includegraphics[width=0.47\textwidth]{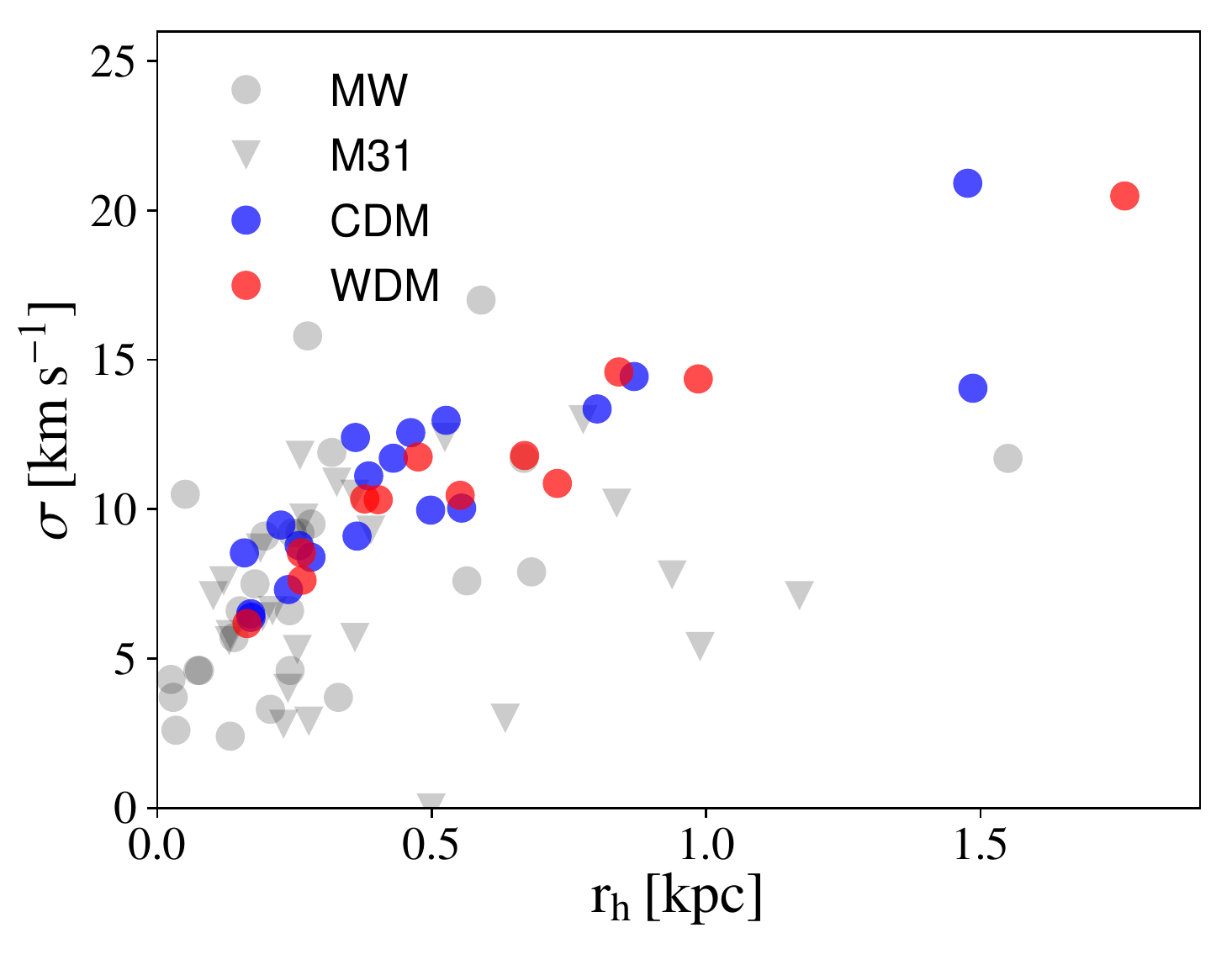}
\vspace{-.35cm}
\caption{Velocity dispersion - size relation. Blue and red circles denote CDM and WDM simulations, respectively, and observations of Milky Way and M31 satellites (for references see section \ref{sec:scaling_relations})
 are shown as grey dots and triangles, respectively.}
\label{fig:field_sigr12}
\end{figure} 

\subsubsection{Scaling relations}\label{sec:scaling_relations}
To compare our simulations with observational data we investigate two observable scaling relations: the  velocity dispersion-size relation and the metallicity-stellar mass relation\footnote{In this and in the following section, we analyzed all simulations containing stars 
except for the WDM run of \textit{g6.31e9}. We removed this simulation from the sample since this galaxy only contains three stellar particles at $z=0$.}. The values of these observables for all simulated galaxies are summarized in Table \ref{tab:sample_observables}.
In Fig. \ref{fig:field_sigr12} we show the one dimensional velocity dispersion (averaged over $x$, $y$ and $z$) and the projected half-mass radi (averaged accordingly).
The data are taken from a compilation from M.Collins (private communication) including data from \citet{Walker2009} for the Milky Way and \citet{Tollerud2012,Tollerud2013,Ho2012, Collins2013, Martin2014} for M31 satellites.
As we already pointed out in \textit{paperI}, due to the absence of tidal effects, the simulations of field galaxies tend to lie on the upper half of the cone spanned by the observational data of satellites. However, WDM and CDM galaxies both occupy the same parameter space and the two populations are not separable.

\begin{figure}
\includegraphics[width=0.47\textwidth]{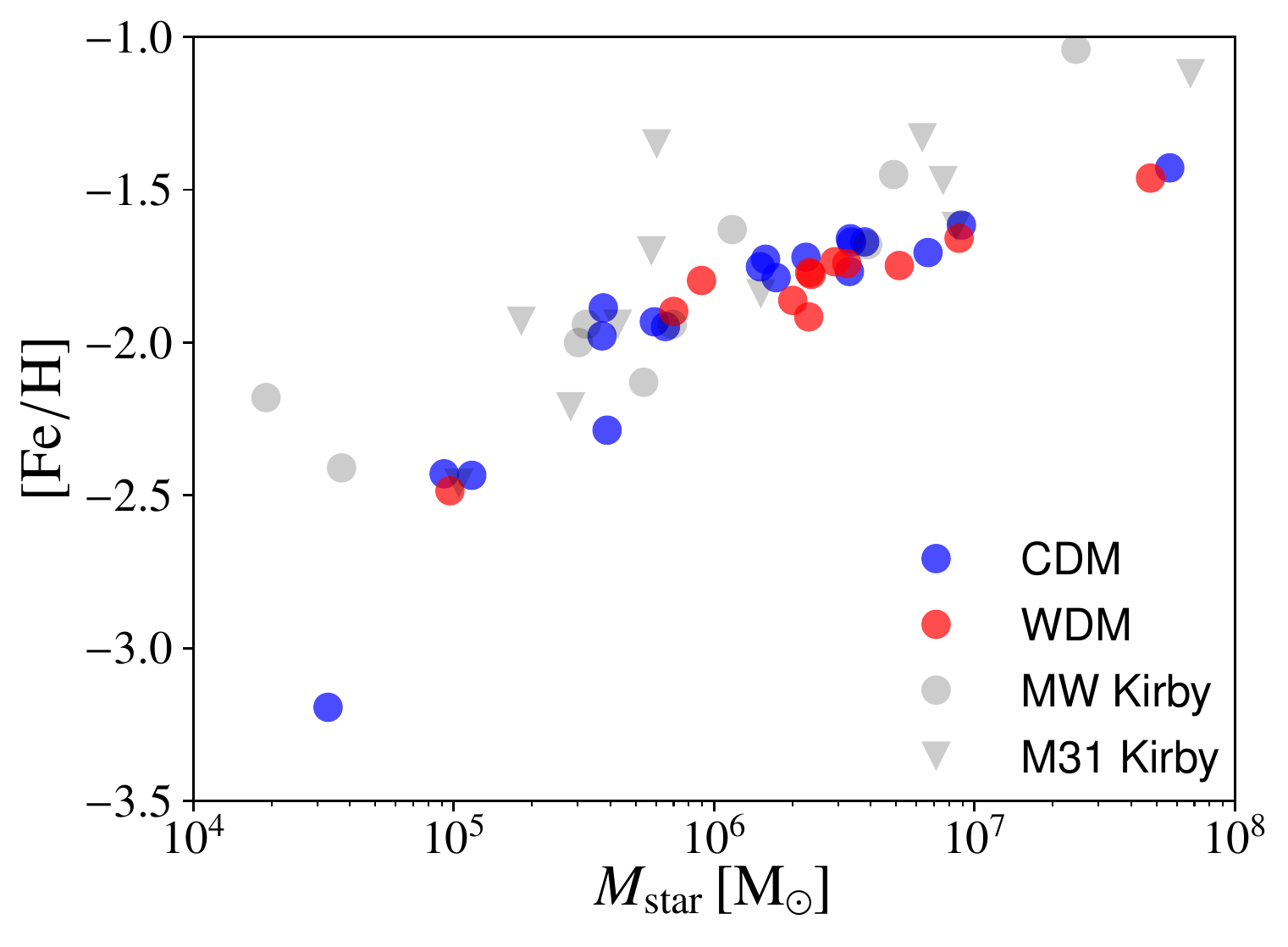}
\vspace{-.35cm}
\caption{Metallicity - stellar mass relation. Blue and red circles denote CDM and WDM simulations, respectively. Observations of Milky Way and M31 satellites from \citet{kirby2014} are shown as grey dots and triangles, respectively.  }
\label{fig:field_metals}
\end{figure}

The same is true for the metallicity-stellar mass relation shown in Fig. \ref{fig:field_metals}. The colour coding is the same as before and observational data of Milky Way and M31 satellites from \citet{kirby2014} are shown as grey dots and triangles, respectively.
Again the CDM simulations as well as the WDM simulations  fulfill the relation down to intermediate masses. For lower stellar masses, however, the star formation in our simulations often happens in a single short burst and time resolution of the gas recycling is too poor to resolve the metal enrichment of the gas, leading to a  lack of metals in galaxies with stellar masses below $ 10^5\,\Msun$ (for more details see \textit{paperI} and \textit{paperII}).

\begin{figure}
\includegraphics[width=0.47\textwidth]{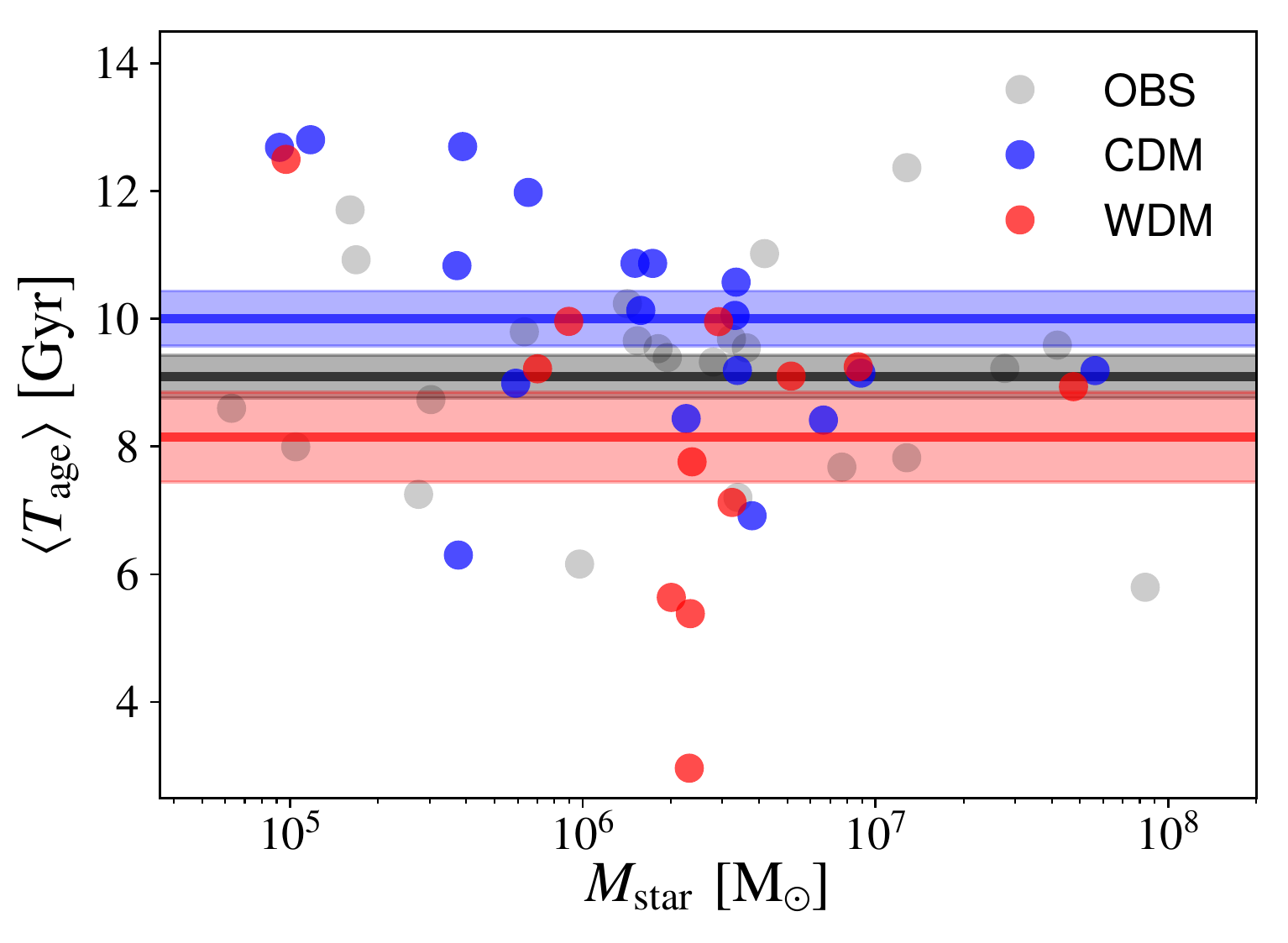}
\vspace{-.35cm}
\caption{Average stellar age as a function of stellar mass. Blue and red circles denote CDM and WDM simulations, respectively, and in the same colour coding lines and shaded areas represent the mean  value and its error.
Observational data (grey) are taken from \citet{Weisz2014}.}
\label{fig:field_age}
\end{figure}

\subsubsection{Stellar ages}\label{sec:stellar_ages}
Since structure formation is  delayed in WDM \citep{Bode2001}, this can leave an imprint in the stellar ages of WDM galaxies \citep{Governato2015, Lovell2017}. In Fig. \ref{fig:field_age} we show the mass averaged stellar age of the galaxies as a function of stellar mass. We additionally show the mean (solid line) and its error (shaded area) for WDM (red), CDM (blue) and the observational data (grey) taken from \citet{Weisz2014}.

CDM galaxies have a mean stellar age of $10.00 \pm 0.43$ Gyr which is slightly larger than the observed value of $9.09\pm0.33$. On the other hand a direct comparison between simulations and observations is quite difficult
since while stellar age is directly recovered from simulations, it is only inferred from observations, and several biases and systematics enter in theobservational derivation of galaxy properties \citep[see for example][]{Guidi2015}.
It is more instructive instead to look at the difference between CDM and WDM at a fixed galaxy formation model. 
WDM galaxies seem to form at later times with respect to CDM  \citep[see also][]{Governato2015, Lovell2017} with a mean age of $8.15\pm0.70 \Gyr$ and seem to struggle to reproduce the observations of galaxies that were in place very early with $\langle T_\mathrm{age} \rangle > 10\Gyr$. In the CDM simulations, those very old galaxies usually form stars before the reionizing background sets in but are quenched during reionization leading to a very old average stellar age. On the other hand the delayed structure formation in WDM makes it very hard for any of these low mass dwarfs to make any stars before reionization and hence strongly reduces the number of very old galaxies. Furthermore WDM predicts very young galaxies where the bulk of star formation happened only $4 \Gyr$ ago. 
This systematic difference in mean stellar ages of satellites and dwarf galaxies might be used in the future to better disentangle CDM and a WDM model with a mass of just $3$ keV.


\subsubsection{Halo structure} 
\begin{figure}
\includegraphics[width=0.47\textwidth]{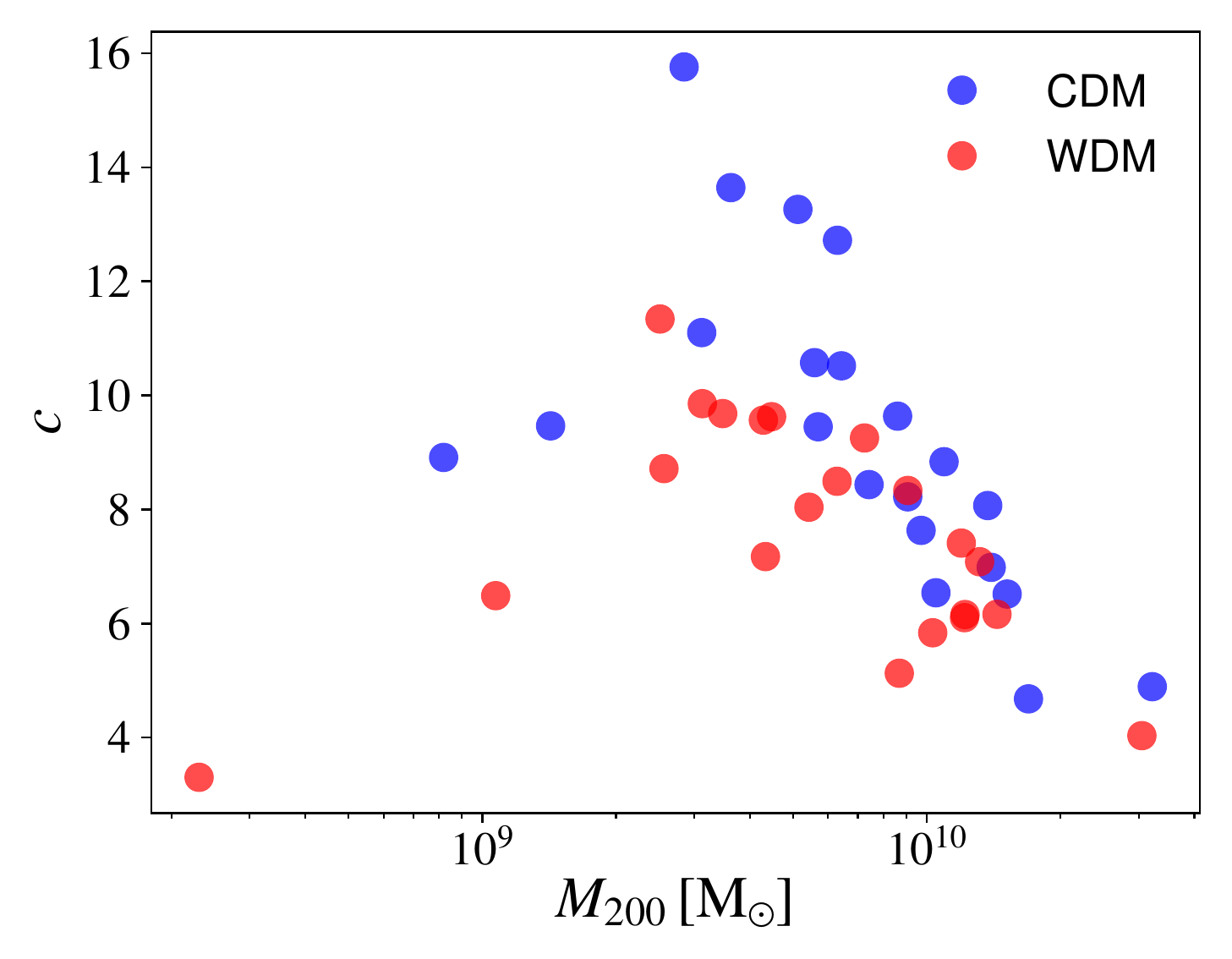}
\vspace{-.35cm}
\caption{The dark matter halo concentrations $c=r_{200}/r_\mathrm{s}$ of the CDM run (blue) and WDM run (red) as a function of the virial mass of the halo. }
\label{fig:field_conc}
\end{figure}

\begin{figure}
\includegraphics[width=0.47\textwidth]{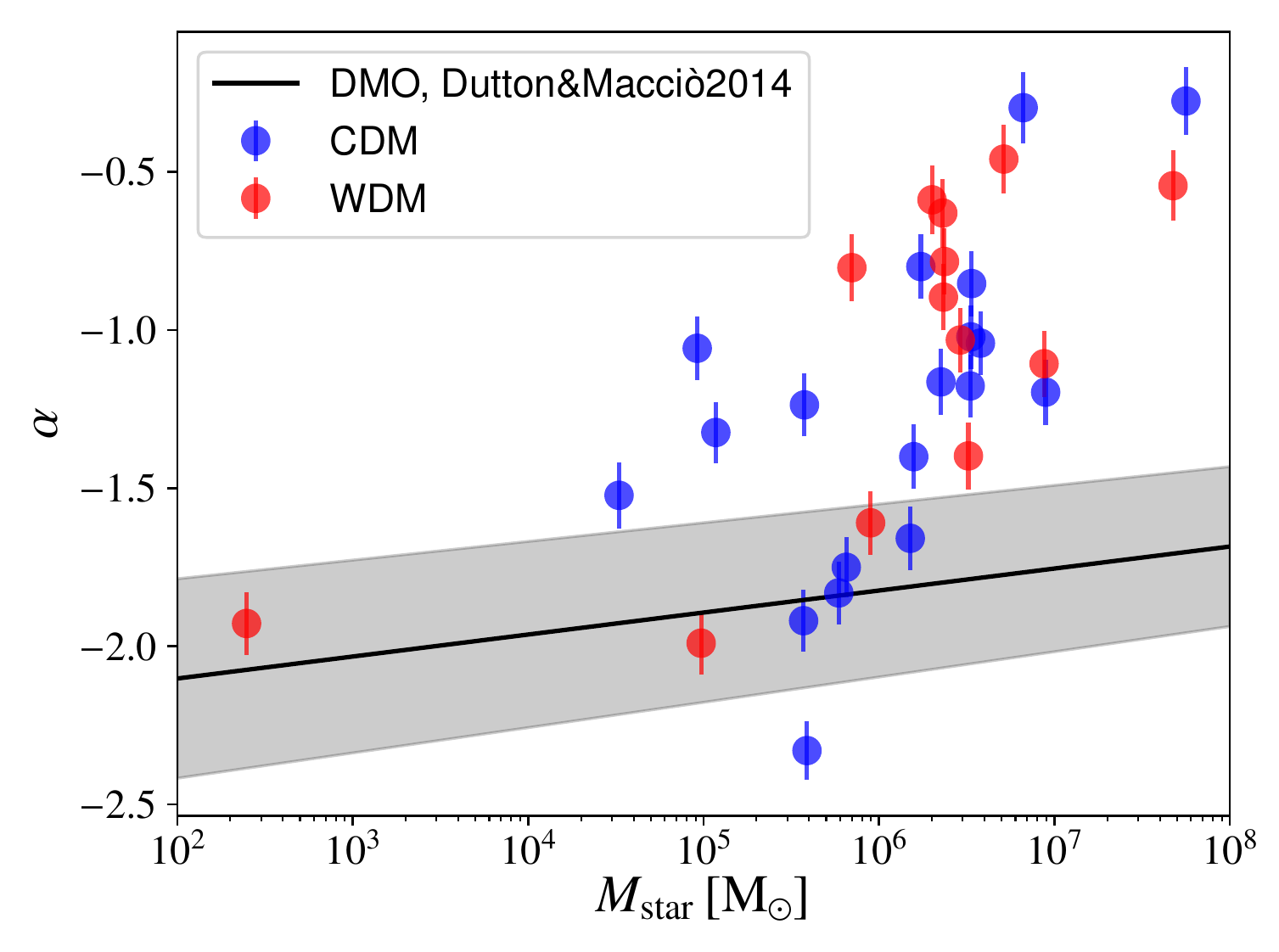}
\vspace{-.35cm}
\caption{The logarithmic inner dark matter density slope $\alpha$ of the CDM (blue) and WDM (red) haloes as a function of stellar mass.
The black line with grey area represent the expected slope and scatter for an Einasto profile based on Nbody results from \citet{Dutton2014}.}
\label{fig:field_alpha}
\end{figure}

Finally we want to investigate the internal dark matter structure of the haloes. The concentrations defined by $c=r_{200}/r_\mathrm{s}$ we show in Fig. \ref{fig:field_conc}, where $r_\mathrm{s}$ is the scale radius of the NFW profile \citep{nfw1996} fit to the dark matter component of the haloes. In agreement with previous studies on WDM
structure formation in N-body simulations \citep{Maccio2012,Schneider2012, Lovell2014} the concentrations of the haloes in the WDM simulations lie below their CDM counterparts.

Another very interesting quantity to look at is the inner logarithmic dark matter density slope $\alpha$. A value of $\alpha =0$ indicates a cored profile  while $\alpha =-1$ is the predicted asymptotic slope for a NFW \citep{nfw1996}. In Fig. \ref{fig:field_alpha} we show the inner logarithmic dark matter density slope, computed between 1 and 2 per cent of the virial radius (see \textit {paperII}) as a function of stellar mass for the CDM runs (blue) and WDM runs (red). 
The black line shows the theoretical expectations for an Einasto  profile obtained extrapolating to low masses the results from pure Nbody 
simulations from \cite{Dutton2014}. 
We use the abundance matching predictions from \cite{Moster2013} to convert between halo mass and stellar mass and the shaded area is the expected
variance on  $\alpha$ assuming a 0.3 dex scatter in the concentration value at a fix halo mass \citep{Dutton2014}. 
As already pointed out in previous work \citep{Pontzen2012,Maccio2012, DiCintio2014a, Madau2014, Chan2015, Dutton2016b,Read2016, Tollet2016,Maccio2017} we see the trend of higher stellar masses producing more and more cored profiles in this mass range. On the other hand, galaxies with little star formation tend to live in very steep dark matter profiles. Also in this case we do not see any particular difference between CDM and WDM simulations, confirming that the main driver of the cusp-core transformation is the star formation efficiency \citep{Dutton2016b} and that the lower concentration of the WDM haloes plays a very minor role.

\subsection{Effects of WDM on tidal stripping of satellites}

\begin{table}
\centering
\caption{Colour coding for the figures \ref{fig:orb_moster} to \ref{fig:orb_alpha}.}

\begin{tabular}{c|c|c|c|c|}

& g4.48e9  &g9.91e9& g1.17e10&g1.23e10 \\ \hline
CDM&\cellcolor{colorl1!70}&\cellcolor{colorl2!70}&\cellcolor{colorl3!70}&\cellcolor{colorl4!70}\\ \hline	
WDM&\cellcolor{colorw1!70}&\cellcolor{colorw2!70}&\cellcolor{colorw3!70}&\cellcolor{colorw4!70}\\ \hline
\end{tabular}
\label{tab:colour}
\end{table}

\subsubsection{Mass loss}
\begin{figure}
\includegraphics[width=0.47\textwidth]{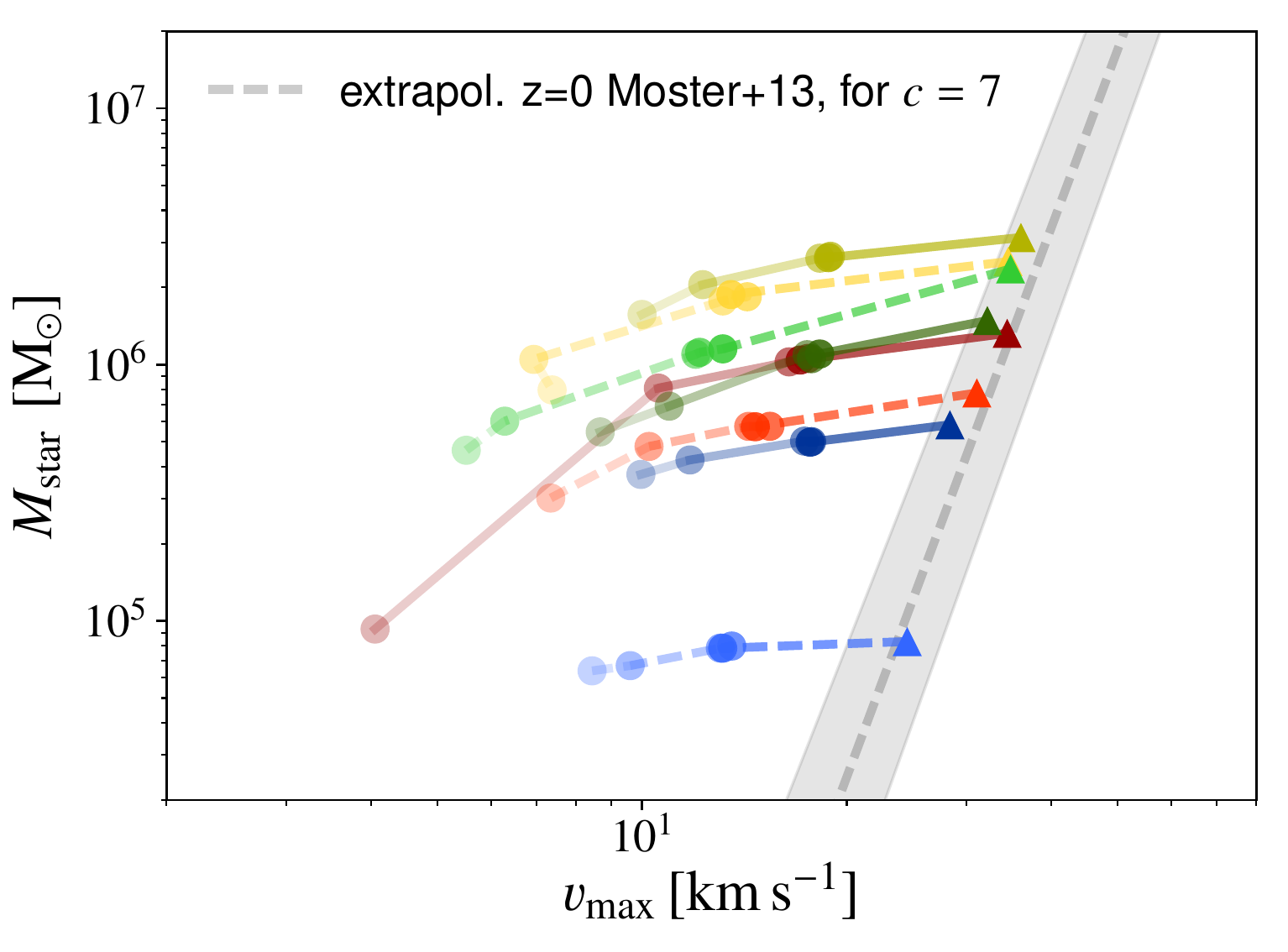}
\vspace{-.35cm}
\caption{The stellar mass within three stellar 3D half-mass radi (measured at the infall time) as a function of the maxium circular velocity. The grey band shows the Moster relation \citep{Moster2013} and its errors translated to a function of $v_\mathrm{max}$ for a concentration of $c=7$. Triangles denote the isolated simulations while the filled circles denote the different orbits. The more violent the orbit, the fainter is the colour of the dots. For the colour coding see Table \ref{tab:colour}.}
\label{fig:orb_moster}
\end{figure}
Now that we know about the effects of WDM on the field dwarfs we want to see if there is a difference in the way galaxies evolve to satellites. To quantify the stellar and dark matter mass loss we show in Fig. \ref{fig:orb_moster} the stellar mass at redshift $z=0$ inside a sphere of three stellar half-mass radi (measured at the infall time) as a function of the maximum value of the circular velocity profile $v_\mathrm{max}$. The colour coding of the plots in this section is explained in Table \ref{tab:colour}. Triangles denote the isolated runs, while the filled circles denote the runs on orbit. We chose a different opacity for each orbit listed in Table \ref{tab:orbits}, where we have ordered the six orbits by their
"disruptiveness", i.e. {\it orbitI} is the most gentle orbit, causing the least deviations from the isolated run (for example in mass-loss) while {\it orbitV} provides the most violent interaction between the satellite and the central object, with the exception of the complete radial infall. 
 While all the galaxies lie on the Moster relation in isolation the WDM counterparts have lower $v_\mathrm{max}$ and stellar mass with the exception of \textit{g1.23e10} that is both more massive and richer in stars. If evolved on a more gentle orbit, stellar masses do not change significantly while $v_\mathrm{max}$ already decreases substantially. On the more violent orbits, however, stellar masses are efficiently reduced. The results are similar to \textit{paperI}. On \textit{orbitV}, however, \textit{g9.91e9} is stripped heavily while its WDM counterpart seems to be more resistant. We will elaborate on this point again in Section \ref{sec:orb_alpha}.

\begin{figure}
\includegraphics[width=0.47\textwidth]{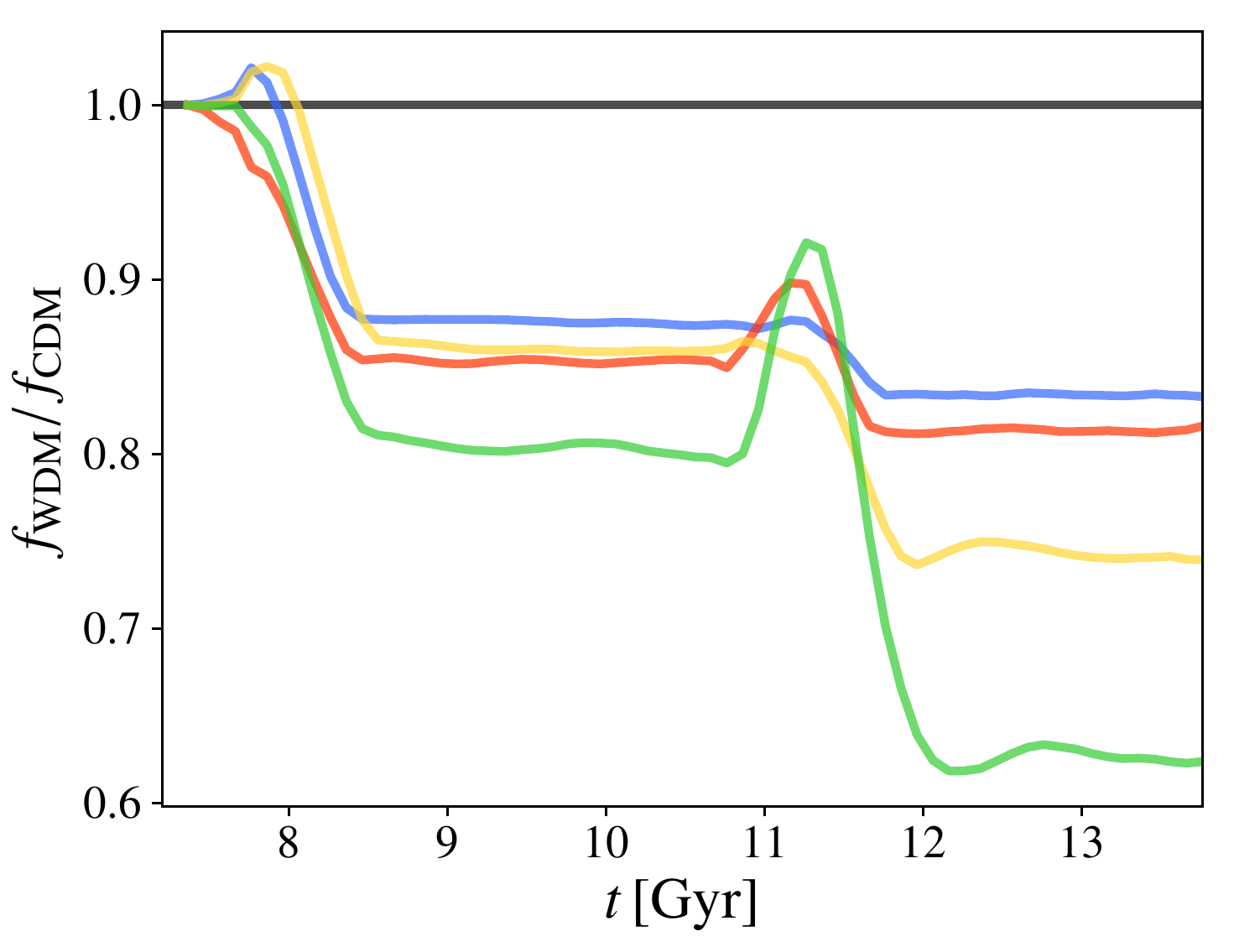}
\vspace{-.35cm}
\caption{Fractional change of maximum circular velocity ($f=v_\mathrm{max}(t)/v_\mathrm{max,infall}$) of the WDM run compared to CDM as a function of time, starting just before the first pericentre passage. All lines are for {\it orbitI}. The black line indicates a ratio of $1$.}
\label{fig:orb_mtime}
\end{figure}

Now we want to compare the mass loss in WDM and CDM directly. Therefore we take a look at the fractional decrease of $v_\mathrm{max}$ given by $f=v_\mathrm{max}(t)/v_\mathrm{max,infall}$. In Fig. \ref{fig:orb_mtime} we show the ratio of the fractional decrease in WDM and CDM over time starting just before the first pericentre passage for {\it orbitI}. After the first encounter we already see the trend that mass at radii of the maximum circular velocity is removed more efficiently in WDM. This trend continues after the second pericentre passage (since we evolve our galaxies in an analytical potential, with no dynamical
friction, the pericentre passage are very similar between WDM and CDM runs). This behaviour can be explained by the lower concentrations of the WDM haloes that we presented in Fig. \ref{fig:field_conc}.
\subsubsection{Central dark matter density slope}\label{sec:orb_alpha}
\begin{figure}
\includegraphics[width=0.47\textwidth]{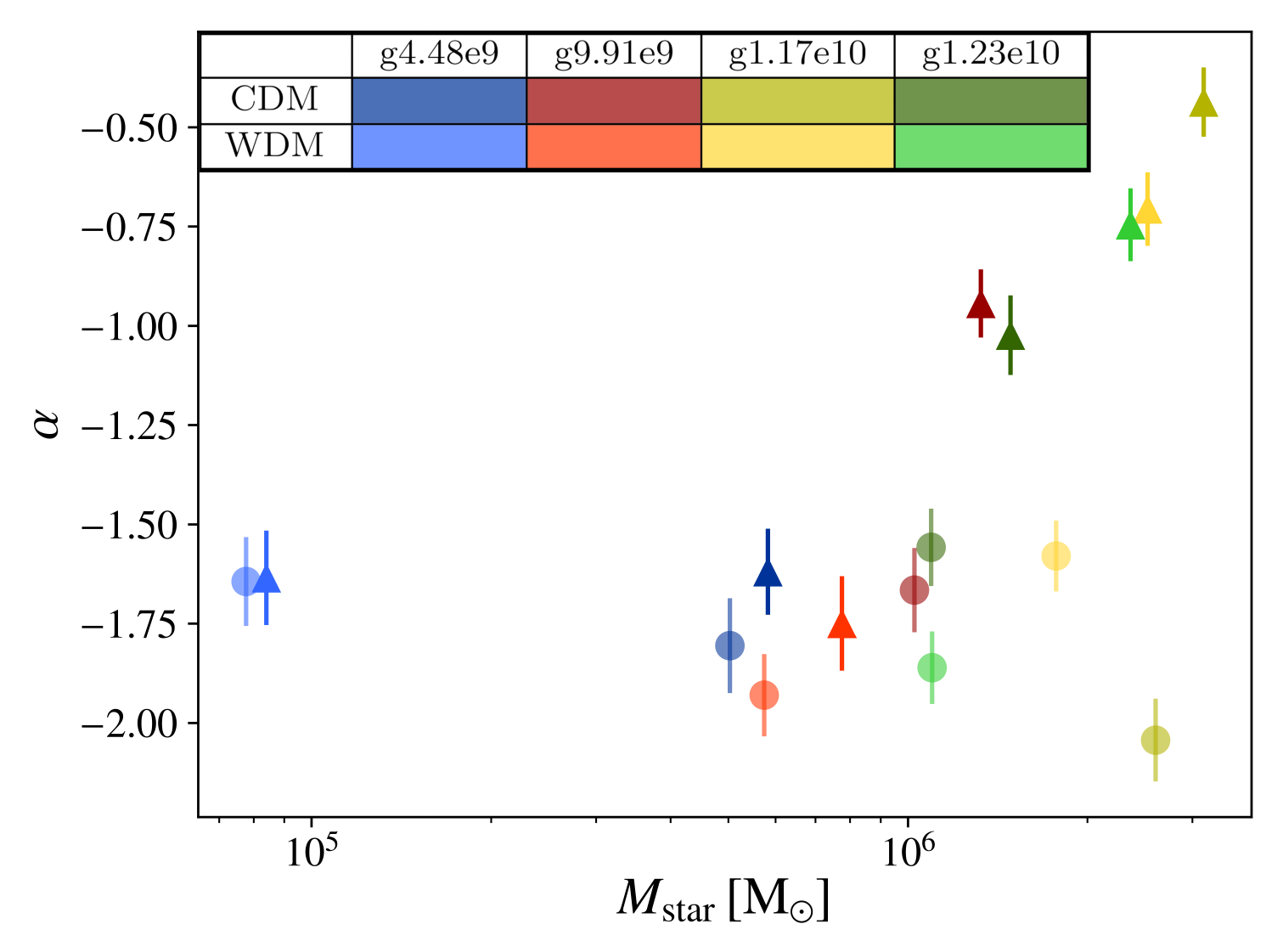}
\vspace{-.35cm}
\caption{Logarithmic inner dark matter density slope $\alpha$ $vs$ stellar mass inside three half-mass radi at infall. Triangles denote the runs in isolation while circles show \textit{orbitI}. The colour coding in the insert is the same as in Table \ref{tab:colour}.}
\label{fig:orb_alpha}
\end{figure}

Since the mass loss and halo survival is linked to the central dark matter structure \citep{Kazantzidis2004,Penarrubia2010,Frings2017} it is interesting to look at the logarithmic inner dark matter density slope $\alpha$. We evaluated $\alpha$ between $1$ and $2$ per cent of the virial radius at infall.
In Fig. \ref{fig:orb_alpha}, we show $\alpha$ versus the stellar mass inside three half-mass radii at infall. The runs in isolation are denoted by the triangles while the filled circles show \textit{orbitI}. For the isolated runs, we apparently see the same behaviour of the profiles becoming more cored for higher stellar masses. As a consequence the density profiles of WDM galaxies seem to be a little steeper as long as they form fewer stars.

However, for \textit{g1.23e10} that is more massive and forms more stars in WDM we end up with a shallower profile. \textit{g9.91e9} shows a remarkable difference in $\alpha$ for WDM and CDM with a value of $\alpha_\mathrm{CDM}$ around $-1$ and $\alpha_\mathrm{WDM}=-1.75$. 
This leaves the central region of \textit{g9.91e9} much more vulnerable to stripping in the CDM case, and also leads to the huge difference in mass loss on \textit{orbitV} (see Fig. \ref{fig:orb_moster}). 
While mass removal is easier in WDM (see Fig. \ref{fig:orb_mtime}) the survivability and stability of the central region and particular of the stars,  is still linked to the dark matter density slope which is usually steeper in WDM
due to the lower star formation. This result might  suggest that WDM haloes might be more resilient than their CDM counterparts at the lowest mass scale. A larger sample and a full cosmological simulations will be needed to fully address this point. 

\section{Discussion and Conclusions}
\label{sec:discussion} 
Warm dark matter with a particle mass above $3\,\mathrm{keV}$ still provides a valid alternative to the standard cold dark matter scenario. In the previous papers of this series \citep{Maccio2017,Frings2017} we have shown that the results of our simulations of field galaxies and satellite galaxies of the Milky Way in CDM are consistent with current observational data. 
In this work we  investigated if the observations constrain the dark matter to be cold or if they allow for a $3\,\mathrm{keV}$ warm dark matter. For that we  presented a sample 42 cosmological simulations of 21 haloes that were run both in CDM and WDM.
As expected from the damping of the power spectrum the virial masses of the WDM haloes lie  below their CDM counterparts for masses below $10^7\,\Msun$. Haloes at larger masses are much less affected.

Like the CDM galaxies, WDM galaxies fulfill the extrapolated abundance matching relation from \citet{Moster2013} even with slightly less scatter. The critical mass for haloes to host luminous galaxies is much higher in WDM, where we find several dark haloes up to a halo mass of $6\times 10^{9}\,\Msun$. Furthermore WDM seems to allow a larger range of masses where one finds both dark and luminous haloes.

We found that WDM and CDM galaxies show no different behaviour in the observable velocity dispersion - size and metallicity - stellar mass relation.
We have further investigated the average stellar formation times and found a possible way to distinguish the two populations: on average the WDM galaxies form slightly later, with a $\langle T_{age} \rangle = 8.15 \pm 0.70$ Gyr ago while CDM galaxies have
earlier formation times with $\langle T_{age} \rangle = 10.00\pm 0.43$.
 This result confirms the predictions from previous studies on the formation of dwarf galaxies in WDM \citep{Governato2015,Chau2017}, even though the current biases and systematic in observationally derived stellar ages \citep{Guidi2015} prevent any firm 
 comparison with observational data.
Finally we found an indication that WDM barely reproduces galaxies at masses around $M_*\approx 10^6\,\Msun$ with mean stellar ages of $12\,\mathrm{Gyr}$, i.e. that formed in the early universe before the UV background sets in. 

In agreement with previous work we find that the WDM haloes are systematically less concentrated than their CDM counterparts. 
On the other hand on small scales, about a few per cent of the virial radius, we confirm that the dark matter density slope is mainly fixed by star formation efficiency \citep[e.g.][]{Tollet2016} and not by the nature of dark matter.
As a consequence at low stellar masses ($10^6\,\Msun$  and below), our (few) data points suggest that the lower star formation efficiency in WDM galaxies  might cause a slightly steeper central slope at those masses, but a larger
sample of simulated galaxies is needed to firmly confirm this effect. 

We used four galaxies of the WDM sample and their CDM counterparts to simulate their evolution in a Milky Way potential as described above.
We witnessed the lower concentrations of the WDM haloes causing the haloes to be more sensitive to mass loss through tidal stripping. Thus their deviation from the stellar mass-halo mass relation is more distinct.

However, the survival of the stellar component of the satellites is also related to central dark matter density slope which for WDM galaxies with a lower star formation efficiency can be much lower. We found that one of the CDM satellites (\textit{g4.48e9}) is stripped close to the point of disruption while its WDM counterpart, 
having a much more cuspy profile, is less affected. Since survival depends on both concentration and central slope, neither WDM nor CDM satellites are a priori  more likely to survive. 

The cusp-core transformation happening in the NIHAO simulations is due to the high density threshold adopted ($\rho_{\rm th} \approx10$ part/cm$^3$) 
as recently studied in details in \cite{Dutton2019} \citep[see also][for an observational evidence for the need of an high star 
formation threshold in cosmological simulations]{Buck2019}. We do not expect this high (observational motivated) threshold for star formation
to affect our results, in fact the star formation rate is very  marginally influenced by the value of  $\rho_{\rm th}$ \citep{Dutton2019}, moreover
the one Gyr time delay seen between star formation in WDM and CDM is much longer than the cooling time at densities between
$0.1<\rho_{\rm th}<10$, which cover the whole spectrum of star formation densities used in current cosmological  simulations by various groups \citep[see][]{Dutton2019, Benitez-Llambay2018}

In good agreement with the CDM studies presented in \textit{paperI} and \textit{paperII}, WDM haloes show an unambiguous steepening of the central dark matter density slope due to stripping.
WDM predicts cuspy density profiles below a stellar mass of $10^6 \Msun$ as well as in
 stripped satellites. An unambiguous observation of a core in these objects would hence force us to deviate from the standard model more drastically than just by changing the dark matter particle mass. 
 
We have shown that current observations of satellites and field galaxies around the Milky Way  struggle to clearly distinguish between CDM and a $3\,\mathrm{keV}$ WDM scenario.
On the other hand future, more precise, observations of the average stellar age might be a promising observable that could help constraining the nature of dark matter.

\section*{Acknowledgements}
The authors  gratefully acknowledge the Gauss Centre for Supercomputing e.V. (www.gauss-centre.eu) 
for funding this project by providing computing time on the GCS Supercomputer SuperMUC at Leibniz Supercomputing Centre (www.lrz.de) and 
the High Performance Computing resources at New York University Abu Dhabi.
Part of this research was also carried out on the {\sc theo} cluster of the Max-Planck-Institut f\"ur Astronomie and on the {\sc hydra} clusters at the Rechenzentrum in Garching.
TB and AVM acknowledge funding from the Deutsche Forschungsgemeinschaft via the SFB 881 program 
``The Milky Way System'' (subproject A1 and A2). JF and AVM acknowledge funding and support through the graduate college {\em Astrophysics of cosmological probes of gravity} by Landesgraduiertenakademie Baden-W{\"u}rttemberg. JF and TB  are members of the International Max-Planck Research School in Heidelberg. AO acknowledges support from the German Science Foundation (DFG) grant 1507011 847150-0.



\bibliographystyle{mnras}
\bibliography{ref} 






\bsp	
\label{lastpage}
\end{document}